\documentstyle[preprint,prd,aps,psfig]{revtex}
\tightenlines
\newcommand{\be}{\begin{eqnarray}}
\newcommand{\ee}{\end{eqnarray}}
\newcommand{\ba}{\begin{array}}
\newcommand{\ea}{\end{array}}
\newcommand{\no}{\nonumber}
\newcommand{\Op}{{\cal O}}

\begin{document}
\draft
\date{\today}
\preprint{
$\ba{r}
\mbox{HUPD-9921}\\
\mbox{KEK-CP-100}
\ea$}
\title{
  Width difference in the $B_s$-$\bar{B}_s$ system
  with lattice NRQCD
  }
\author{
  S. Hashimoto$^{\mbox{\scriptsize\ a}}$,
  K-I. Ishikawa$^{\mbox{\scriptsize\ a}}$,
  T. Onogi$^{\mbox{\scriptsize\ b}}$,
  and
  N. Yamada$^{\mbox{\scriptsize\ b}}$
}
\address{
  $^{\mbox{\scriptsize\ a}}$
  High Energy Accelerator Research Organization (KEK),
  Tsukuba 305-0801, Japan\\
  $^{\mbox{\scriptsize\ b}}$
  Department of Physics, Hiroshima University,
  Higashi-Hiroshima 739-8526, Japan
}
\maketitle
\begin{abstract}
  We present a lattice calculation of the $B_s-\bar{B}_s$
  transition matrix element through a four-quark operator 
  $\Op_S$ = $\bar{b}(1-\gamma_5) s\ \bar{b} (1-\gamma_5) s$,
  which gives a leading contribution in the calculation of
  the width difference $\Delta\Gamma_s$ in the $1/m_b$
  expansion.
  The NRQCD formulation is used to describe $b$ quark on the
  lattice. 
  Using the next-to-leading formula of 
  Beneke \textit{et al.}, we obtain
  $(\Delta\Gamma/\Gamma)_s$ = 0.151(37)(45)(17),
  where the first error reflects the uncertainty of the
  $B_s$ meson decay constant, the second error comes from our
  calculation of the matrix element of $\Op_S$, and the
  third represents unknown $1/m_b$ correction.
\end{abstract}
\pacs{PACS number(s): 12.38.Gc, 12.39.Hg, 13.20.He, 14.40.Nd}

\section{Introduction}
\label{sec:Introduction}

The mixing and decays of the $B_s^0-\bar{B}_s^0$ system play
a complementary role to the $B^{\pm}$ and $B_d^0-\bar{B}_d^0$
systems in studying the flavor mixing and CP violation
\cite{Buras_Fleischer_HF2}.
In particular, if the width difference of the
$B_s^0-\bar{B}_s^0$ system is sufficiently large,
the angle $\phi_3$($\gamma$) of the unitarity triangle can be measured 
through untagged modes such as 
$B_s\rightarrow D_s^{(*)} K^{(*)}$ or 
$B_s\rightarrow D^{*}\phi$ 
\cite{Dunietz_95,Fleischer_Dunietz_96-97},
which would be promising not only because the method is
theoretically clean but also feasible at future hadron colliders.

The width difference $\Delta\Gamma_{B_s}$ of the
$B_s-\bar{B}_s$  systems is calculated most reliably using
the heavy quark expansion \cite{Beneke_Buchalla_Dunietz_96},
and the size of a ratio $(\Delta\Gamma/\Gamma)_{B_s}$ is
roughly estimated as $(\Delta\Gamma/\Gamma)_{B_s}$ =
0.16($^{+0.11}_{-0.09}$).
Now that the perturbative error has been reduced by the recent 
calculation of  the next-to-leading order (NLO) QCD corrections 
\cite{Beneke_et_al_99}, the largest remaining uncertainty comes
from the matrix elements  
$\langle\bar{B}_s|\Op_X(\mu_b)|B_s\rangle$ 
($X=L$ or $S$) of four-quark operators
\begin{eqnarray}
  \label{eq:def_of_O}
  \Op_L &=& \bar{b}\gamma_\mu (1-\gamma_5) s\ 
            \bar{b}\gamma_\mu (1-\gamma_5) s,\\
  \Op_S &=& \bar{b} (1-\gamma_5) s\ \bar{b} (1-\gamma_5) s.
\end{eqnarray}

Lattice QCD is one of the most suitable tools for the nonperturbative
computation of matrix elements such as the decay constants and the 
bag parameters. In fact a number of extensive studies,
including ours \cite{Yamada_et_al_99}, have already been
done to obtain  $B_L$\footnote{
  We use the notation $B_L$ instead of $B_B$ to explicitly
  indicate that it represents a matrix element of $\Op_L$.}
\cite{Hashimoto_lat99},
which is a matrix element of the former operator 
$\Op_L$ normalized by its vacuum saturation approximation.
On the other hand, the matrix element $B_S$ 
for the latter operator $\Op_S$ has been calculated in 
Ref.~\cite{Gupta_Bhattacharya_Sharpe_97} only for the 
heavy-light meson around charm quark mass regime. 
It is required to perform a thorough study of $B_S$ in order 
to give a reliable prediction of the $B_s$ width difference. 
The matrix element of $\Op_S$ is also required in the
evaluation of the amplitude $\Delta M$ of the
$B_{(s)}-\bar{B}_{(s)}$ mixing,
if we assume the physics beyond the Standard Model
such as the supersymmetric models~\cite{BS_in_SUSY}.

In this paper, we present a quenched lattice calculation of
the matrix element of $\Op_S$ using the NRQCD formalism~\cite{NRQCD}
for heavy quark and the $O(a)$-improved Wilson action~\cite{SW}
for light quark.
The NRQCD formalism is formulated as an inverse heavy quark
mass expansion, and our action and operators consistently
include entire $O(p/m_Q)$ terms,
where $p$ denotes a typical spatial momentum of a heavy quark
inside a heavy-light meson.
Higher order contribution of $O(p^2/m_Q^2)$ is also studied by
introducing all necessary terms, and we find those effect is
small for the $b$ quark mass.

In this work one-loop matching of the operator $\Op_S$ between continuum
and lattice regularizations is performed in the limit of infinitely
heavy quark mass,
so that the systematic error of $O(\alpha_s/(am_Q))$
is not removed.
Since the $b$ quark mass in the lattice unit is not
extremely large, $O(\alpha_s/(am_Q))$ gives a non-negligible effect
in our final result, which could be as large as about 10\% in a naive
order counting argument.

Using the NLO formula of Ref.~\cite{Beneke_et_al_99} and our
results for the matrix elements of $\Op_S$ as well as of
$\Op_L$~\cite{Yamada_et_al_99}, we obtain a prediction 
$(\Delta\Gamma/\Gamma)_{B_s}$ = 0.151(37)(45)(17).
The first error originates in the $B_s$ meson decay
constant $f_{B_s}$ = 245(30) MeV~\cite{Hashimoto_lat99} used
to normalize the 
matrix elements, and the second is from our calculation of
the matrix element of $\Op_S$.
The error from $B_L$ is negligible, since it gives only a
small contribution to the width difference.
The last error is a crude estimate of the $O(1/m_b)$
correction as discussed in Refs.
\cite{Beneke_Buchalla_Dunietz_96,Beneke_et_al_99}.

This paper is organized as follows.
We briefly summarize the NLO formula of Ref.~\cite{Beneke_et_al_99}
for the width difference in the next section.
We present the perturbative matching of the operator $\Op_S$
in Section~\ref{sec:Operator_matching}, while the detail of
the one-loop calculation is given in Appendix.
We describe our simulation methods in Section~\ref{sec:Simulations},
and our results for the matrix element and the width difference are
given in Section~\ref{sec:Results}.
In section~\ref{sec:uncertainties},
we attempt to estimate the size of $O(\alpha_s/(am_Q))$ error,
which is specific to our work with NRQCD.
Section~\ref{sec:Discussion} is devoted to a comparison of
our result with a previous work by Gupta \textit{et al.}
\cite{Gupta_Bhattacharya_Sharpe_97},
who obtained the same matrix element using the relativistic
lattice action around charm quark mass.
Finally our conclusion is given in Section~\ref{sec:Conclusion}.
A preliminary report of this work is included in
Ref.~\cite{Yamada_lat99}.

\section{Width difference of $B_s$ mesons}
\label{sec:Width_difference_of_B_s_mesons}

In this section we briefly summarize the formula to
give the width difference of $B_s$ mesons, which was
obtained by Beneke \textit{et al.} in Ref.~\cite{Beneke_et_al_99}.

The width difference in the $B_s-\bar{B}_s$ system is given by
\begin{equation}
  \label{eq:width_difference_formula_1}
  \Delta\Gamma_{B_s} = - 2 \frac{1}{2M_{B_s}}
  \langle\bar{B}_s| \mbox{Im\ } i \int d^4x
  T {\cal H}_{eff}(x){\cal H}_{eff}(0) |B_s\rangle,
\end{equation}
where ${\cal H}_{eff}$ is a $\Delta B$=1 weak transition
Hamiltonian.
The main contribution comes from a transition
$b\bar{s}\rightarrow c\bar{c}$ followed by
$c\bar{c}\rightarrow \bar{b}s$, and other contributions
mediated by penguin operators are also included
\cite{Beneke_et_al_99}.

Using the $1/m_b$ expansion, the transition operator
$\mbox{Im\ } i \int d^4x 
T {\cal H}_{eff}(x){\cal H}_{eff}(0)$
is represented by the local four-quark operators $\Op_L$ and 
$\Op_S$, 
which leads to the following 
formula at the next-to-leading order \cite{Beneke_et_al_99},
\begin{eqnarray}
  \label{eq:width_difference_formula_2}
  \left(\frac{\Delta\Gamma}{\Gamma}\right)_{B_s}
  & = & 
  \frac{16\pi^2 B(B_s\rightarrow Xe\nu)}{
    g(z)\tilde{\eta}_{QCD}}
  \frac{f_{B_s}^2 M_{B_s}}{m_b^3} |V_{cs}|^2
  \nonumber\\
  & & \times
  \left( G(z) \frac{8}{3} B_L(m_b)
    + G_S(z) \frac{5}{3} \frac{B_S(m_b)}{{\cal R}(m_b)^2}
    + \sqrt{1-4z} \delta_{1/m}
  \right).
\end{eqnarray}
Here, the quantity $B(B_s\rightarrow Xe\nu)$ is
the semi-leptonic decay branching ratio.
The factors $g(z)=1-8z+8z^3-z^4-12z^2\ln z$ ($z=m_c^2/m_b^2$)
and $\tilde{\eta}_{QCD}$ represent the phase space factor and
the QCD correction, respectively.
The coefficients
$G(z)$ and $G_S(z)$ are functions including the
next-to-leading QCD corrections, and their numerical values
are given in Table 1 of Ref.~\cite{Beneke_et_al_99}.

$B_L(m_b)$ and $B_S(m_b)$ are the $B$ parameters defined
with the $\overline{MS}$ scheme at the renormalization scale
$\mu_b=m_b$. 
Their definitions are
\begin{eqnarray}
  \label{eq:B_L_definition}
  B_L(\mu_b) & \equiv &
  \frac{
    \langle\bar{B}_s| \Op_L(\mu_b) |B_s\rangle}{
    \frac{8}{3} \langle\bar{B}_s| A_0 |0\rangle
                \langle 0| A_0 |B_s\rangle}
  =
  \frac{
    \langle\bar{B}_s| \Op_L(\mu_b) |B_s\rangle}{
    \frac{8}{3}f_{B_s}^2 M_{B_s}^2}, \\
  \label{eq:B_S_definition}
  B_S(\mu_b) & \equiv &
  \frac{
    \langle\bar{B}_s| \Op_S(\mu_b) |B_s\rangle}{
    \frac{5}{3} \langle\bar{B}_s| P(\mu_b) |0\rangle
                 \langle 0| P(\mu_b) |B_s\rangle}
  =
  \frac{
    \langle\bar{B}_s| \Op_S(\mu_b) |B_s\rangle}{
    -\frac{5}{3}f_{B_s}^2 M_{B_s}^2}
  \times {\cal R}(\mu_b)^2.
\label{eq:R2}
\end{eqnarray}
In the last expression in Eq.~(\ref{eq:R2}),
we change the normalization of
$\langle\bar{B}_s|\Op_S(\mu_b)|B_s\rangle$
with the decay constant $f_{B_s}$ by factoring out the ratio
\begin{equation}
  \label{eq:R=A/P}
  {\cal R}(\mu_b) \equiv 
  \left| \frac{\langle 0| A_0 |B_s\rangle}{
               \langle 0| P(\mu_b) |B_s\rangle} \right|.
\end{equation}
Using the equation of motion the ratio
${\cal R}(\mu_b)$ is expressed in terms of the quark masses
$m_b$ and $m_s$ as
\begin{equation}
  \label{eq:R_by_quark_mass}
  {\cal R}(\mu_b) = 
  \frac{\bar{m}_b(\mu_b)+\bar{m}_s(\mu_b)}{M_{B_s}},
\end{equation}
where $\bar{m}_b(\mu_b)$ and $\bar{m}_s(\mu_b)$ denote the
quark masses defined with the $\overline{MS}$ scheme at
scale $\mu_b$.

Finally $\delta_{1/m}$ denotes $1/m_b$ corrections, which
may be estimated using the factorization approximation
\cite{Beneke_Buchalla_Dunietz_96}.

Numerically evaluating the coefficients in the right hand side of
Eq.~(\ref{eq:width_difference_formula_2}), we obtain 
\begin{equation}
  \label{eq:width_difference_formula_3}
  \left(\frac{\Delta\Gamma}{\Gamma}\right)_{B_s}
  = \left(\frac{f_{B_s}}{245 {\rm MeV}} \right)^2
  \left[ 0.008 B_L(m_b) 
    + 0.150 \frac{B_S(m_b)}{{\cal R}(m_b)^2} - 0.086
  \right],
\end{equation}
where we choose a recent world average of unquenched lattice
simulations $f_{B_s}$ = 245(30) MeV for the central value of 
the decay constant \cite{Hashimoto_lat99}.
In the following sections we present a calculation of the
parameter $B_S(m_b)/{\cal R}(m_b)^2$.
Our calculation of $B_L(m_b)$ is already available
in Ref.~\cite{Yamada_et_al_99}.

\section{Operator matching}
\label{sec:Operator_matching}

In this section, we present the perturbative matching of
continuum operator $\Op_S$ to the corresponding operators
defined on the lattice. 
We follow the calculation method in Ref.
\cite{Ishikawa_Onogi_Yamada_99}, where the one-loop matching of the
operator $\Op_L$ is presented.

Following the definition in Ref.~\cite{Beneke_et_al_99},
we adopt modified minimal subtraction ($\overline{MS}$)
with the Naive Dimensional Regularization (NDR) scheme
for the continuum operator $\Op_S(\mu_b)$,
in which $\gamma_5$ anticommutes with all $\gamma_{\mu}$'s.
The subtraction of evanescent operators is done with the
definition given by Eqs.~(13)-(15) of Ref.~\cite{Beneke_et_al_99}.
The renormalization scale $\mu_b$ is set to the $b$ quark mass $m_b$. 

While in the numerical simulations we apply the NRQCD formalism
\cite{NRQCD} to the heavy quarks, in the perturbative calculation the
heavy quarks are treated as a static quark~\cite{Eichten_Hill_90}.
More comments on this approximation will be given in the end of this
section.
The light quarks and gauge fields are described by the $O(a)$-improved
SW quark action~\cite{SW} and the standard Wilson (plaquette) action,
respectively, in both of the perturbative calculation and the
numerical simulations.

The operators involved in the calculation are
\begin{eqnarray}
  \label{eq:O_S}
  \Op_S &=& \bar{b}^i P_L s^i\ \bar{b}^j P_L s^j,
  \\
  \label{eq:O_S_tilde}
  \tilde{\Op}_S
        &=& \bar{b}^i P_L s^j\ \bar{b}^j P_L s^i,             \\
  \Op_L &=& \bar{b}^i\gamma_\mu P_L s^i\
            \bar{b}^j\gamma_\mu P_L s^j,                      \\
  \Op_P &=& 2\ \bar{b}^i\gamma_\mu P_L s^i\
               \bar{b}^j\gamma_\mu P_R s^j
         + 4N\ \bar{b}^i P_L s^i\ \bar{b}^j P_R s^j,       \\
  \Op_R &=& \bar{b}^i\gamma_\mu P_R s^i\
            \bar{b}^j\gamma_\mu P_R s^j,                      \\
  \Op_{SD} &=& \bar{b}^i P_L (\vec{\gamma}\cdot a\vec{D}) s^i\
               \bar{b}^j P_L s^j,
  \label{eq:O_SD}                   \\
  \Op_{LD} &=& \bar{b}^i\gamma_\mu P_L (\vec{\gamma}\cdot a\vec{D}) s^i\
               \bar{b}^j\gamma_\mu P_L s^j,
  \label{eq:O_LD}         \\
  \Op_{PD} &=& 2\ \bar{b}^i\gamma_\mu P_R (\vec{\gamma}\cdot a\vec{D}) s^i\
                  \bar{b}^j\gamma_\mu P_L s^j
            + 4N\ \bar{b}^i P_R (\vec{\gamma}\cdot a\vec{D}) s^i\
                 \bar{b}^j P_L s^j,
  \label{eq:O_PD}
\end{eqnarray}
where $P_L$ and $P_R$ are chirality projection operators
$P_{L/R}=1\mp\gamma_5$.
Color indices $i$ and $j$ run from one to $N$ for SU($N$)
gauge theory and $a$ denotes lattice spacing.
In the continuum, in which the chiral symmetry for light quark is
preserved, the operator $\Op_S$ mixes only with $\tilde{\Op}_S$ and
$\Op_L$.
On the lattice, however, the chiral symmetry is explicitly broken with
the SW action, so that additional operators with opposite chirality,
$\Op_P$ and $\Op_R$, appear in the operator matching.

Other operators $\Op_{SD}$, $\Op_{LD}$ and $\Op_{PD}$ are higher
dimensional operators introduced to cancel a discretization error of
$O(\alpha_s a)$.
However we neglect this discretization error in the numerical
simulations.
The result of $O(a\alpha_s)$ matching coefficients is presented only
for future use.

Here we show the one-loop result of the matching.
We leave the detail of the calculations for
Appendix~\ref{sec:Appendix}.
The continuum operator $\Op_S(\mu_b)$ is expressed by lattice
operators $\Op_X^{\rm lat}(1/a)$ as follows~\cite{Ishikawa_lat99},
\begin{eqnarray}
  \Op_S(\mu_b)
  &=& \Bigg[\ 1 + \frac{\alpha_s}{4 \pi}
        \Bigg\{ \frac{4}{3}\ln\left(a^2 m_b^2\right)
                + \frac{16}{3}\ln\left(\frac{\mu_b^2}{m_b^2}\right)
                - 3.86
        \Bigg\}
      \Bigg] \Op_S^{\rm lat}(1/a) \nonumber\\
   &&+ \frac{\alpha_s}{4 \pi}
       \Bigg[\ - \frac{2}{3}\ln\left(a^2 m_b^2\right)
            + \frac{1}{3}\ln\left(\frac{\mu_b^2}{m_b^2}\right)
            + 3.91
       \Bigg] \Op_L^{\rm lat}(1/a)  \nonumber\\
   &&+ \frac{\alpha_s}{4 \pi}\ \Big[\ 0.77\ \Big]
       \Op_P^{\rm lat}(1/a)
     + \frac{\alpha_s}{4 \pi}\ \Big[\ 0.13\ \Big]
       \Op_R^{\rm lat}(1/a) \nonumber\\
   &&+ \frac{\alpha_s}{4 \pi}\ \Big[\ -6.88\ \Big]
       \Op_{SD}^{\rm lat}(1/a)
     + \frac{\alpha_s}{4 \pi}\ \Big[\  2.58\ \Big]
       \Op_{LD}^{\rm lat}(1/a)
     + \frac{\alpha_s}{4 \pi}\ \Big[\  1.15\ \Big]
       \Op_{PD}^{\rm lat}(1/a).
  \label{eq:O_S_matching}
\end{eqnarray}
The operator $\tilde{\Op}_S$ is eliminated from the right hand
side using a identity
$\tilde{\Op}_S = -\Op_S - \frac{1}{2} \Op_L$, 
which is valid up to $O(p/m_Q)$.

The heavy-light axial vector current $A_0$ is also necessary
to normalize the matrix element.
The one-loop matching of $A_0$ is already known as
\cite{Borrelli_Pittori_92,Morningstar_Shigemitsu_98,Ishikawa_Onogi_Yamada_99}
\begin{eqnarray}
  A_0
  &=& Z_A(1/a) A_0^{\rm lat}(1/a)
    + Z_{A_D}(1/a) A_{D0}^{\rm lat}(1/a)    \no\\
  &=& \Bigg[ 1 + \frac{\alpha_s}{4\pi}
        \left[\ 2\ln(a^2 m_b^2) - 16.561\
        \right]\
      \Bigg]\ A_0^{\rm lat}(1/a)
     - \frac{\alpha_s}{4\pi}[\ 13.01\ ]\ A_{D0}^{\rm lat}(1/a)
  \label{eq:axial_matching},
\end{eqnarray}
where $A_0$ and $A_{D0}$ are defined as
\begin{eqnarray}
  A_0    &=& \bar{b}\gamma_0\gamma_5 s,\\
  A_{D0} &=& \bar{b}\gamma_0\gamma_5 (\vec{\gamma}\cdot a\vec{D}) s.
\end{eqnarray}
The higher dimensional operator $A_{D0}$ is introduced to
remove the $O(\alpha_s a)$ errors.

In Eqs.~(\ref{eq:O_S_matching}) and (\ref{eq:axial_matching}),
we apply the tadpole improvement~\cite{Lepage_Mackenzie_93}
using $u_0=1/8\kappa_c$ as an average link variable. 
The normalization of the light quark filed is
$\sqrt{1-3\kappa/4\kappa_c}$. 

To obtain the matching coefficient for $B_S/{\cal R}^2$ we 
combine Eqs.~(\ref{eq:O_S_matching}) and (\ref{eq:axial_matching}),
and linearize the perturbative expansion in $\alpha_s$. 
Omitting the higher dimensional operators, which we
neglect in the following numerical simulations, we obtain 
\begin{eqnarray}
  \label{eq:B_S_matching}
  B_S(\mu_b)/{\cal R}(\mu_b)^2 & = &
     \Bigg[\ 1 + \frac{\alpha_s}{4 \pi}
       \Bigg\{ -\frac{8}{3}\ln\left(a^2 m_b^2\right)
                + \frac{16}{3}\ln\left(\frac{\mu_b^2}{m_b^2}\right)
                + 29.26
       \Bigg\}
      \Bigg] \hat{B}_S^{\rm lat} \nonumber\\
   &&+ \frac{\alpha_s}{4 \pi}
       \Bigg[\ - \frac{2}{3}\ln\left(a^2 m_b^2\right)
            + \frac{1}{3}\ln\left(\frac{\mu_b^2}{m_b^2}\right)
            + 3.91
       \Bigg] \hat{B}_L^{\rm lat}  \nonumber\\
   &&+ \frac{\alpha_s}{4 \pi}\ \Big[\ 0.77\ \Big]
       \hat{B}_P^{\rm lat}
     + \frac{\alpha_s}{4 \pi}\ \Big[\ 0.13\ \Big]
       \hat{B}_R^{\rm lat},
\end{eqnarray}
where $\hat{B}_X^{\rm lat}$ ($X$ = $S$, $L$, $P$, or $R$)
are `$B$ parameters' defined by 
\begin{equation}
  \hat{B}_X^{\rm lat} =
  \frac{\langle{\cal O}_X^{\rm lat}(1/a)\rangle}{
    -\frac{5}{3} \langle A_0^{\rm lat}(1/a) \rangle^2},
  \label{eq:lat_B}
\end{equation}
which we measure in the numerical simulations.

Before closing this section,
we should clarify the remaining uncertainty arising from the static
approximation in the matching coefficients.
In the simulation, the heavy quarks are described by the NRQCD action
including the $O(p/m_Q)$ or $O(p^2/m_Q^2)$ corrections consistently.
The $b$ quark field, which constitutes the operators measured in the
simulation, is also improved through the same order as the action by
the inverse Foldy-Wouthuysen-Tani transformation $R^{-1}$ as
\begin{eqnarray}
 b & =&  R^{-1}  \left( \begin{array}{c}
                          Q\\
                  \chi^{\dagger}
                         \end{array}
                         \right), \no
\end{eqnarray}
where $Q$ and $\chi^{\dagger}$ are the two-component quark and
anti-quark fields in the NRQCD action.
Therefore the truncation error only starts from $O(p^2/m_Q^2)$ or
$O(p^3/m_Q^3)$, which depends on the accuracy of our action and
operators, even at the tree level matching.
On the other hand, the static approximation in the perturbative
calculation only leads to a lack of finite mass effects in the
matching coefficients, but does not change the truncation error.
Therefore, using the matching coefficients derived in this section
the result has the $O(\alpha_s/(a m_Q))$ error.

\section{Simulations}
\label{sec:Simulations}

The numerical simulations to extract
$\hat{B}_X^{\rm lat}$ are almost the same as in our previous
paper \cite{Yamada_et_al_99}, in which we calculated $B_L$. 
We carried out a quenched simulation on 250 $16^3 \times 48$
lattices at $\beta$=5.9.
The inverse lattice spacing from the string tension is 1.64 GeV.
We employ the SW action for light quark~\cite{SW} with
mean field improved $c_{sw}=1/{u_0}^3$ with $u_0=0.8734$.
The heavy quark is treated by two sets of NRQCD actions
and fields~\cite{NRQCD} as was done in Ref.~\cite{Yamada_et_al_99}:
one is truncated at $O(p/m_Q)$ and
the other includes entire $O(p^2/m_Q^2)$ corrections.
We use the difference between the results from these sets to estimate
the size of truncation error of the $p/m_Q$ expansion.

For the strong coupling constant used in the perturbative
matching, we choose the $V$-scheme coupling $\alpha_V(q^*)$ 
with $q^*=1/a$, $2/a$ or $\pi/a$.
Their numerical values are $\alpha_V(1/a)$ = 0.270,
$\alpha_V(2/a)$=0.193 and $\alpha_V(\pi/a)$ = 0.164. 

Other details of our simulations, such as the exact
definition of the NRQCD action and the mass parameters used, 
are found in the previous paper \cite{Yamada_et_al_99}.

\section{Results}
\label{sec:Results}

\begin{figure}
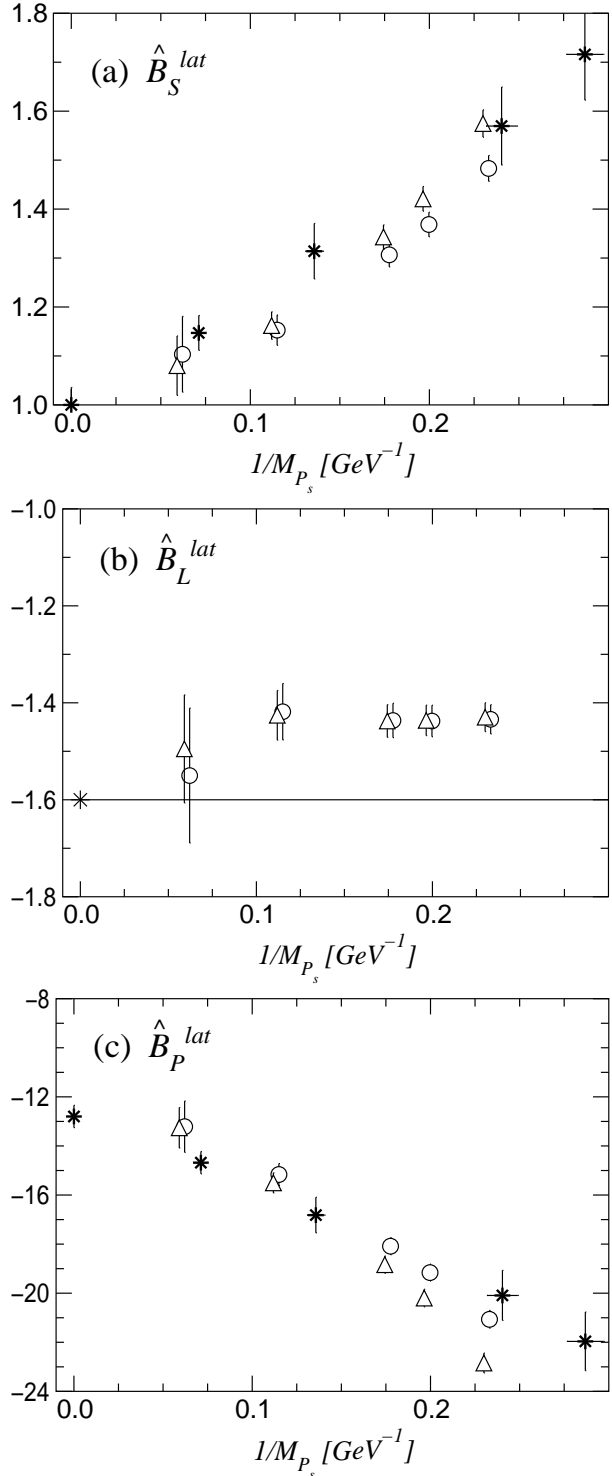

  \begin{center}
    \leavevmode\psfig{file=mdep.BS.eps,width=8cm,clip=}\\
    \leavevmode\psfig{file=mdep.BL.eps,width=8cm,clip=}\\
    \leavevmode\psfig{file=mdep.BP.eps,width=8cm,clip=}\\
    \caption{$1/M_{P_s}$ dependence of 
      (a) $\hat{B}_S^{\rm lat}$, 
      (b) $\hat{B}_L^{\rm lat}$ (=$\hat{B}_R^{\rm lat}$),
      and
      (c) $\hat{B}_P^{\rm lat}$.
      The results with $O(p/m_Q)$ accuracy (circles) are
      compared to those with
      $O(p^2/m_Q^2)$ (triangles) accuracy.
      The vacuum saturation approximation is shown by
      crosses.}
    \label{fig:bx}
  \end{center}
\end{figure}

Figure \ref{fig:bx} shows the mass dependence of
$\hat{B}^{\rm lat}_X$ ($X$ = $S$, $L$ or $P$) defined in
Eq.~(\ref{eq:lat_B}).
$\hat{B}^{\rm lat}_R$ is equal to $\hat{B}^{\rm lat}_L$
because of a symmetry under parity transformation.
The light quark mass is interpolated to the strange quark mass.
Since the light quark mass dependence is very small,
in the following analysis
we do not consider the error arising from the interpolation.
The inverse heavy-light meson mass $1/M_{P_s}$, for which
the light quark mass is also interpolated to the strange
quark mass, is used as a horizontal axis.

The difference between two results with different accuracies
of the $p/m_Q$ expansion does not exceed a few percent at
the $b$ quark mass, as explicitly presented in the figure by
different symbols: circles for $O(p/m_Q)$ and triangles for
$O(p^2/m_Q^2)$ accuracy.
It justifies the use of the nonrelativistic expansion for
the $b$ quark.

As we pointed out in the previous paper~\cite{Yamada_et_al_99},
the vacuum saturation approximation (VSA) gives a good approximation
of the lattice data.
In the static limit, it becomes $\hat{B}_S^{(VSA)}$=1,
$\hat{B}_L^{(VSA)}$ = $-8/5$ and
$\hat{B}_P^{(VSA)}$ = $-64/5$. 
For the finite heavy quark mass, the axial current and the
pseudoscalar density involved in the VSA have different
matrix elements.
As a result, a mass dependence appears in the VSA of
$\hat{B}_X$, as plotted by crosses (a flat line for
$\hat{B}_L$) in Fig.~\ref{fig:bx}.
It is remarkable that the VSA explains the $1/M_{P_s}$
dependence of the data very nicely.

\begin{figure}
  \begin{center}
    \leavevmode\psfig{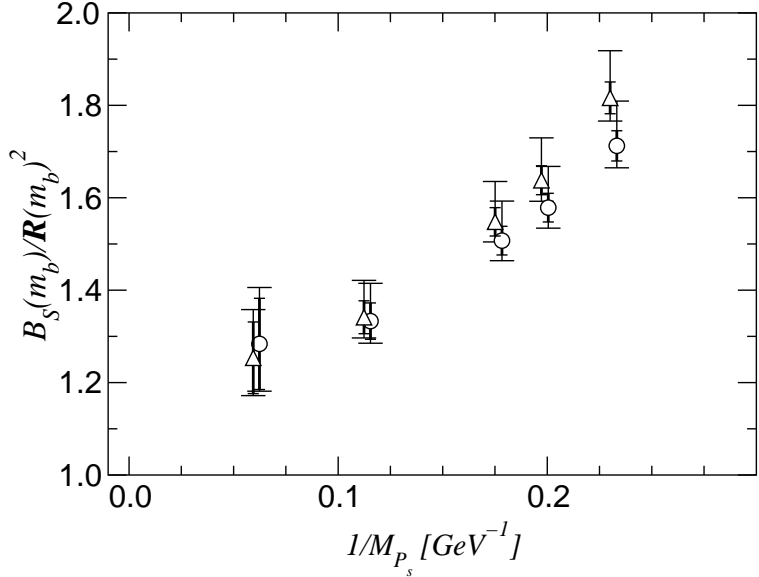}
    \caption{$1/M_{P_s}$ dependence of
      $B_S(m_b)/{\cal R}(m_b)^2$.
      Results with the $O(p/m_Q)$ (circles) and
      $O(p^/m_Q^2)$ (triangles) accuracies are shown.
      The smaller error bars represent statistical errors,
      while the uncertainties, obtained from a quadratic sum of
      the statistical uncertainty and difference between
      the central values with $\alpha_V(1/a)$ and $\alpha_V(\pi/a)$,
      are shown by the larger error bars.
      The central values are obtained with $\alpha_V(2/a)$.
      }
    \label{fig:BsoverR}
  \end{center}
\end{figure}

We combine the results for $\hat{B}^{\rm lat}_X$ to obtain
$B_S(\mu_b)/{\cal R}(\mu_b)^2$ using Eq.~(\ref{eq:B_S_matching}).
The renormalization scale $\mu_b$ is set to the $b$ quark
pole mass $m_b$ = 4.8 GeV according to Ref.~\cite{Beneke_et_al_99}.  
Figure \ref{fig:BsoverR} presents the $1/M_{P_s}$ dependence
of $B_S(m_b)/{\cal R}(m_b)^2$ obtained with the $O(p/m_Q)$ (circles)
and $O(p^2/m_Q^2)$ (triangles) accuracies and
using $\alpha_V(2/a)$=0.193 as a coupling constant
in the perturbative matching.
Typical size of the perturbative error may be evaluated by
comparing the results obtained with different coupling constants. 
For this purpose, we also calculate the results with
$\alpha_V(\pi/a)$=0.164 and $\alpha_V(1/a)$=0.270,
which are considered in the larger error bars
in Fig.~\ref{fig:BsoverR}.
We find that they give at most 5\% differences at the $b$ quark mass.

Our numerical results interpolated to the physical $B_s$
meson mass $M_{B_s}$ = 5.37 GeV are
\begin{itemize}
\item for $O(p/m_Q)$ accuracy
  \begin{equation}
    \frac{B_S(m_b)}{{\cal R}(m_b)^2}
    = \left\{ 
      \begin{array}{lll}
        1.51(3) & \mbox{\ at\ } & q^* = \pi/a\\
        1.54(3) & \mbox{\ at\ } & q^* = 2/a\\
        1.61(3) & \mbox{\ at\ } & q^* = 1/a
      \end{array} \right.,
  \end{equation}
\item for $O(p^2/m_Q^2)$ accuracy
  \begin{equation}
    \frac{B_S(m_b)}{{\cal R}(m_b)^2}
    = \left\{ 
      \begin{array}{lll} 
        1.56(3) & \mbox{\ at\ } & q^* = \pi/a\\
        1.59(3) & \mbox{\ at\ } & q^* = 2/a\\
        1.67(3) & \mbox{\ at\ } & q^* = 1/a
      \end{array} \right.,
  \end{equation}
\end{itemize}
where the error represents the statistical error. The variation
due to the choice of the coupling constant $\alpha_V(q^*)$ is
explicitly shown.

We attempt to estimate the size of systematic uncertainty in
our result using an order counting of missing contributions. 
As we found in the previous paper \cite{Yamada_et_al_99},
the dominant uncertainties are
\be
&& O(\alpha_s/(am_Q))        \sim 15\%, \no\\
&& O(\alpha_s^2)             \sim 10\%, \no \\
&& O(a^2\Lambda_{QCD}^2)     \sim
   O(a\Lambda_{QCD}\alpha_s) \sim 5\%,
\no
\ee
when we assume $\Lambda_{QCD}\sim$ 300 MeV and $\alpha_s\sim$ 0.3.
Although a naive order counting yields
$O(\alpha_s/(am_Q))\sim$ 10\%,
we take more conservative estimate $\sim$ 15\%, which is
suggested in the study of bilinear operators as we will
discuss in the next section.
The effect of the truncation of the nonrelativistic expansion is
negligible as we explicitly see in the difference between
the two simulations of the $O(p/m_Q)$ and $O(p^2/m_Q^2)$
accuracies. 

We finally obtain
\begin{equation}
  \label{eq:B_S_value}
  \frac{B_S(m_b)}{{\cal R}(m_b)^2} = 1.54(3)(30),
\end{equation}
where the first error represents the statistical error,
while the second is obtained by adding the sources of
systematic uncertainty in quadrature. 

Using this result and the result for $B_L(m_b)$ previously
obtained in Ref.~\cite{Yamada_et_al_99},
 $B_L(m_b)$ = 0.75(2)(12), we find
\begin{equation}
  \label{eq:result}
  \left(\frac{\Delta\Gamma}{\Gamma}\right)_{B_s}
  = 0.151(37)(45)(17),
\end{equation}
from Eq.~(\ref{eq:width_difference_formula_3}).
The first error comes from the uncertainty in the decay
constant $f_{B_s}$ = 245(30) MeV, which is taken from the
current world average of unquenched lattice calculations
\cite{Hashimoto_lat99}.
The second reflects the error in the calculation of
$B_S/{\cal R}^2$ presented above, and the last is obtained
by assuming that the size of error in the $1/m_b$ correction
$\delta_{1/m}$ in Eq.~(\ref{eq:width_difference_formula_2})
is $\pm$ 20\%.
The current experimental bound is
$(\Delta\Gamma/\Gamma)_{B_s} <$ 0.42 \cite{DELPHI_BS_99}.

The central value of our result in Eq.~(\ref{eq:result}) is much
larger than the estimate 0.054$^{+0.016}_{-0.032}$ obtained
by Beneke \textit{et al.}~\cite{Beneke_et_al_99}. 
The main reasons are
\begin{itemize}
\item The unquenched lattice result of $f_{B_s}$ is about 15--20\%
  larger than the previously known quenched result.
\item The central value of our result for $B_S/{\cal R}^2$
  is larger than the previous 
  value obtained from the relativistic lattice calculation
  \cite{Gupta_Bhattacharya_Sharpe_97}, which is used in
  Ref.~\cite{Beneke_et_al_99}.
  We will compare our result with theirs in
  section~\ref{sec:Discussion}.
\end{itemize}

\section{finite mass effects in the matching coefficients}
\label{sec:uncertainties}

In this section, we attempt to estimate the size of
$O(\alpha_s/(am_Q))$ error arising from the lack of
necessary one-loop correction, by taking the ratio
${\cal R}(m_b)$ defined in Eq.~(\ref{eq:R=A/P}) as an example.
Although the $O(\alpha_s/(am_Q))$ errors in bilinear
operators and in the bag parameters are independent,
it would still be useful to explicitly see
the size of the error in a quantity, for which 
the correct one-loop coefficient is known.

\begin{figure}
  \begin{center}
    \leavevmode\psfig{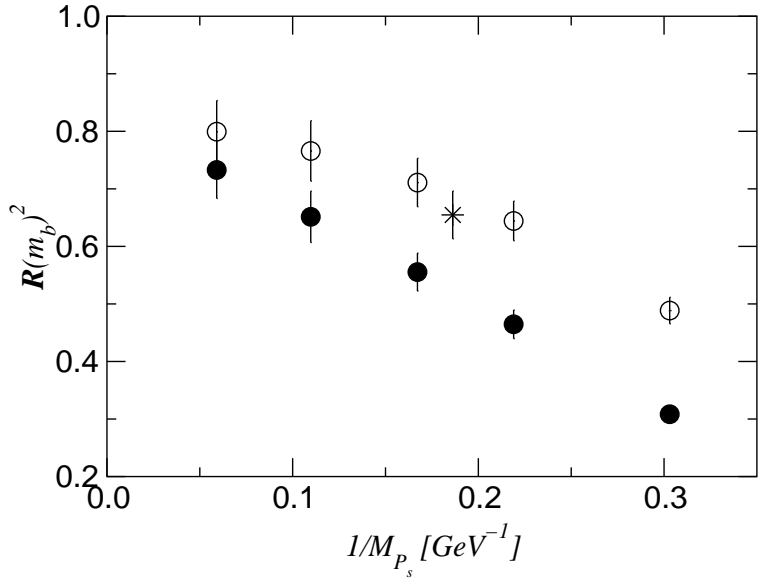}
    \caption{$1/M_{P_s}$ dependence of ${\cal R}(m_b)^2$
      evaluated with the method 1 (circles) and 2 (star).
      See the text for the detail. 
      Open and filled symbols are obtained with and without
      the $1/(am_Q)$ corrections in the one-loop coefficients. }
    \label{fig:R2}
  \end{center}
\end{figure}

We compare the values of ${\cal R}(m_b)^2$ obtained with
the following methods.
\begin{enumerate}
\item 
  The quantity ${\cal R}(m_b)^2$ may be explicitly
  calculated in lattice simulation by measuring the 
  matrix elements of axial-vector and pseudoscalar density.
  Results of the JLQCD collaboration obtained with the NRQCD
  action~\cite{JLQCD_99} are plotted in
  Fig.~\ref{fig:R2} as a function of $1/M_{P_s}$.
  One-loop matching to the continuum operator are calculated
  for two different lattice actions: static (filled circles)
  and NRQCD (open circles)~\cite{Sakamoto_et_al}.
\item
  The equation of motion may be used to obtain
  \begin{equation}
    {\cal R}(m_b)^2 =
    \left(\frac{\bar{m}_b(m_b)+\bar{m}_s(m_b)}{M_{B_s}}
    \right)^2.
  \end{equation}
  For the phenomenological values 
  $\bar{m}_b(m_b)$=4.1--4.4~GeV and 
  $\bar{m}_s(2 {\rm GeV})$=0.06--0.17 GeV~\cite{PDG}, 
  which corresponds to $\bar{m}_s(m_b)$=0.05--0.14 GeV,
  we obtain ${\cal R}(m_b)^2$=0.66(5), which is shown by a
  star in Fig.~\ref{fig:R2}.
\end{enumerate}
The data obtained with the correct NRQCD matching coefficients
(open circles) show a nice agreement with the
phenomenological estimate (star).
This suggests that the error in the calculation of the matrix
element with correct matching coefficient is under good control.
On the other hand, the data with the static matching
coefficients (filled circles) are significantly lower,
indicating a large systematic errors of
$O(\alpha_s/(am_Q))$. 
The difference of ${\cal R}(m_b)^2$ between the two matching
calculations is around 15\% for the $B_s$ meson mass.
We use this number for the estimation of the systematic
error of $O(\alpha/(am_Q))$ for $B_S(m_b)/{\cal R}(m_b)^2$
in Sec.~\ref{sec:Results}.

\section{Discussion}
\label{sec:Discussion}

It is instructive
to compare our result with the previous lattice calculation
by Gupta, Bhattacharya and
Sharpe~\cite{Gupta_Bhattacharya_Sharpe_97}, 
who used the Wilson fermion action for heavy quark with the
mass around charm quark.
Conversion of their result to the definition used in this
paper is given in Ref.~\cite{Beneke_et_al_99}, which yields
$B_S(2.33~{\rm GeV})$=0.81 and
$\tilde{B}_S(2.33~{\rm GeV})$=0.87.
A $B$-parameter for the operator $\tilde{\Op}_S$
(\ref{eq:O_S_tilde}) is denoted as $\tilde{B}_S$.
With the renormalization group evolution, it becomes
$B_S(m_b)$=0.75 and $\tilde{B}_S(m_b)$=0.85 at $\mu_b=m_b$. 
The error was not quoted except for the statistical one,
which is 0.01 for each quantity.
In order to compare the results obtained with different
heavy quark mass, it is
necessary to remove a logarithmic dependence on the heavy
quark mass.
We, therefore, define $\Phi_{B_S}(m_b)$ as
\begin{equation}
  \label{eq:phi_s}
  \Phi_{B_S}(m_b)
  = \Bigg[\ 1 - 2 \frac{\alpha_s(m_b)}{4\pi}
                    \ln\bigg(\frac{m_Q^2}{m_b^2}\bigg)\
      \Bigg]\ B_S(m_b)
  + \frac{2}{5}\frac{\alpha_s(m_b)}{4\pi}
  \ln\bigg(\frac{m_Q^2}{m_b^2}\bigg)\
  \tilde{B}_S(m_b),
\end{equation}
where $m_Q$ denotes the heavy quark mass used in the
simulation.
In the calculation of Gupta \textit{et al.}
\cite{Gupta_Bhattacharya_Sharpe_97} it is about the charm
quark mass $m_Q$ = $m_c$ = 1.4~GeV.
Using the coupling constant $\alpha_s(m_b)$= 0.22 corresponding to
$\Lambda_{\overline{MS}}^{(4)}$=0.327~GeV and
${\cal R}(m_b)^2$ obtained with the method 2 in the
previous section, we obtain 
\be
\Phi_{B_S}^{\rm GBS}(m_b)/{\cal R}(m_b)^2 = 1.20,
\label{eq:Phi_over_R2_GBS}
\ee
which may be compared with our result of $B_S(m_b)/{\cal R}(m_b)^2$
in Eq.~(\ref{eq:B_S_value}).

The central value of our result is significantly higher than 
Eq.~(\ref{eq:Phi_over_R2_GBS}), which is one of the reasons
of our larger value of
$\left(\frac{\Delta\Gamma}{\Gamma}\right)_{B_s}$
compared to that of Ref.~\cite{Beneke_et_al_99}.
We note, however, that the calculation with the unimproved
relativistic action could suffer from large $O(am_Q)$
error, which is not even estimated in
Ref.~\cite{Gupta_Bhattacharya_Sharpe_97}.
In our NRQCD calculation, on the other hand, all possible
systematic uncertainties are considered, but unfortunately
the large systematic error of $O(alpha_s/am_Q)$ is left to
be removed.
Thus, at this stage we conclude that that the present accuracy of both
calculations is not enough for a detailed comparison.

\section{Conclusion}
\label{sec:Conclusion}

The width difference $\Delta\Gamma_s$ in the $B_s-\bar{B}_s$
mixing is expressed by the matrix elements of local
four-quark operators in the $1/m_b$ expansion.
The operator $\Op_S$ gives a dominant contribution among
them and the nonperturbative calculation of its matrix
element is essential for a reliable calculation of the width
difference
\cite{Beneke_Buchalla_Dunietz_96,Beneke_et_al_99}.
We calculated a parameter $B_S(m_b)/{\cal R}(m_b)^2$,
which is the matrix element normalized with a square of the $B_s$
meson decay constant as defined in Eq.~(\ref{eq:B_S_definition}), 
using lattice NRQCD formalism for heavy quark.

From a quenched simulation at $\beta$=5.9
with the $O(a)$-improved light quark action,
we obtain $B_S(m_b)/{\cal R}(m_b)^2$ = 1.54(3)(30),
where statistical and systematic errors are given in that order.
By explicitly performing two calculations with the different
accuracies, we found that the $O(p^2/m_Q^2)$ corrections
in the NRQCD action and operators is only a few percent.
One of the dominant sources of the systematic error is
a lack of one-loop matching coefficients with finite mass corrections.
We used the one-loop coefficients for the static action
instead, which introduces a systematic error of order
$\alpha_s/(am_Q)\sim$ 15\%.

The large remaining uncertainty in our final result for
$(\Delta\Gamma/\Gamma)_s$, Eq.~(\ref{eq:result}),
comes partly from the error in our calculation of
$B_S(m_b)/{\cal R}(m_b)^2$.
Another important source is present in the $B_s$ meson decay
constant $f_{B_s}$, as it appears as $f_{B_s}^2$ in the formula. 

We also discussed a comparison of our result with the previous one.
We found that the central value of our result is significantly larger.
However, since both calculations suffer from large systematic
uncertainties, it would be fair to say that the discrepancy
between the two results is not significant at the present level.

\section*{Acknowledgment}

Numerical calculations have been done on Paragon XP/S at
Institute for Numerical Simulations and Applied Mathematics
in Hiroshima University.
We are grateful to S. Hioki and H. Matsufuru for allowing us to use
their program.
S.H an T.O. are supported by the Grants-in-Aid
of the Ministry of Education (Nos. 11740162, 10740125).
K-I.I. would like to thank the JSPS
for Young Scientists for a research fellowship.

\appendix
\section{}
\label{sec:Appendix}

A matrix element of the continuum operator $\Op_S$ with free 
quark external states is expressed at one-loop order as
\begin{eqnarray}
  \label{eq:O_S_cont}
  \langle\Op_S(\mu)\rangle
  &=& \Bigg[\ 1 + \frac{\alpha_s}{4 \pi}
                \Bigg\{   \frac{13N^2 - 18N + 9}{4N}
                        + \frac{- 3N^2 + 2N + 5}{2N}
                          \ln\left(\frac{\lambda^2}{m_b^2}\right) \no\\
  & & \hspace{55mm}        + \frac{3N^2 - 4N - 1}{N}
                          \ln\left(\frac{\mu^2}{m_b^2}\right)
                \Bigg\}
      \Bigg]\langle \Op_S \rangle_0
  \nonumber\\
  & & + \frac{\alpha_s}{4 \pi}
    \Bigg[ - \frac{11N - 9}{2N}
           - \frac{N + 1}{N}
             \ln\left(\frac{\lambda^2}{m_b^2}\right)
           - \frac{2(N - 2)}{N}
             \ln\left(\frac{\mu^2}{m_b^2}\right)
    \Bigg]\langle \tilde{\Op}_S \rangle_0
  \nonumber\\
  & & - \frac{\alpha_s}{4 \pi} \frac{3(N-1)}{4N}
      \langle \Op_L \rangle_0   \no\\
  & &+ \frac{\alpha_s}{4\pi}
    \left[-\frac{2\pi}{3}\frac{1}{a\lambda}\right]
    \langle \Op_{PD}\rangle_{0},
\end{eqnarray}
where
$\langle \Op_X \rangle_0$ denotes a tree level matrix
element of operator $\Op_X$, and the gluon mass $\lambda$ is
introduced to regularize the infrared divergence.
The evanescent operators are subtracted according to
Eqs.(13)-(15) of Ref.~\cite{Beneke_et_al_99}.
The expression is expanded in $1/m_b$ and only the leading
terms are written.

The corresponding expression for the lattice operator is
\cite{DiPierro_Sachrajda_98,Gimenez_Reyes_99}
\begin{eqnarray}
  \langle \Op_S^{\rm lat}(1/a) \rangle
  & = & \Bigg[\ 1 + \frac{\alpha_s}{4 \pi}
        \Bigg\{   \frac{ - 3N^2 + 2N + 5}{2N}\ln\left(a^2\lambda^2\right)
              + \frac{N^2 - 1}{2N}\left(f + f^I + e^{(R)} + u_0^{(2)}
                                  \right) 
  \nonumber\\
  & & \hspace{22mm}
              + \frac{ N^2 - 2}{N} d_1
              - \frac{1}{2N}c         
              + \frac{2N - 1}{6N}(v+v^I)
              - \frac{N + 1}{3N}J_1
      \Bigg\}\
    \Bigg]\langle \Op_S \rangle_0    
  \nonumber\\
  & & + \frac{\alpha_s}{4 \pi}
    \Bigg[ - \frac{N + 1}{N}\ln\left(a^2\lambda^2\right)
           + d_1 + \frac{1}{2}c + \frac{N - 2}{6N}(v+v^I)
           + \frac{N + 1}{3N}J_1
    \Bigg]\langle \tilde{\Op}_S \rangle_0    
  \nonumber\\
  & &+ \frac{\alpha_s}{4 \pi}\ \frac{1}{4}\ \Big[ d_2 - d^I \Big]
    \langle \Op_P \rangle_0
  \nonumber\\
  & &+ \frac{\alpha_s}{4 \pi}\ \frac{N-1}{2N}\ \Big[ w + w^I \Big]
    \langle \Op_R \rangle_0
  \nonumber\\
  & &+ \frac{\alpha_{s}}{4\pi}\frac{(N+1)(N-2)}{N}
    \left[-(1-c_{\rm sw})\ln\left(a^{2}\lambda^{2}\right)
          -\left(V+V^{I}\right)\right]
    \langle \Op_{SD}\rangle_{0}
  \nonumber\\
  & &+ \frac{\alpha_{s}}{4\pi}\frac{1}{2}
    \left[\ (1-c_{\rm sw})\ln\left(a^{2}\lambda^{2}\right)
          + \left(V+V^{I}\right)\right]
    \langle \Op_{LD}\rangle_{0}
  \nonumber\\
  & &+ \frac{\alpha_{s}}{4\pi}
    \left[- \frac{2\pi}{3}\frac{1}{a\lambda}-\frac{1}{4}
            \left( U+U^{I}\right)\right]
    \langle \Op_{PD}\rangle_{0},
\end{eqnarray}
where the constants
$c$, $d_1$, $d_2$, $e^{(R)}$, $f$, $v$, $w$
$d^I$, $f^I$, $v^I$, $w^I$ $U$, $U^I$, $V$, $V^I$ and $J_1$
are defined in Ref.
\cite{Eichten_Hill_90,Flynn_Hernandez_Hill_91,DiPierro_Sachrajda_98,Gimenez_Reyes_99,Ishikawa_Onogi_Yamada_99}
and their numerical values are tabulated in
Table~\ref{tab:PERT}. 
The coefficients with the superscript $I$ denotes the terms 
appearing with the $O(a)$-improvement.
$u_0^{(2)}$ comes from the tadpole improvement of
the light quark wave function renormalization, and
is also given in Table \ref{tab:PERT}.

Matching above results and using a Fierz relation
$\langle \tilde{\Op}_S \rangle_0
= - \langle \Op_S \rangle_0 - \frac{1}{2} \langle \Op_L \rangle_0$,
which is satisfied in the static limit,
we obtain for $N$=3
\begin{eqnarray}
  \Op_S(\mu)
  &=& \Bigg[\ 1 + \frac{\alpha_s}{4 \pi}
          \Bigg\{\  10 + \frac{4}{3}\ln\left(a^2 m_b^2\right)
                  + \frac{16}{3}\ln\left(\frac{\mu^2}{m_b^2}\right)
                  - \frac{4}{3}\left(f+f^I+e^{(R)}+u_0^{(2)}\right)
  \nonumber\\&&\hspace{64mm}
                  - \frac{4}{3}d_1 + \frac{2}{3}c
                  - \frac{2}{9}( v + v^I )
                  + \frac{8}{9}J_1\
          \Bigg\}\
    \Bigg] \Op_S^{\rm lat}(1/a)
  \nonumber\\
  & & + \frac{\alpha_s}{4 \pi}
    \Bigg[\   \frac{3}{2} - \frac{2}{3}\ln\left(a^2 m_b^2\right)
            + \frac{1}{3}\ln\left(\frac{\mu^2}{m_b^2}\right)
            + \frac{1}{2}d_1 + \frac{1}{4}c + \frac{1}{36}(v+v^I)
            + \frac{2}{9}J_1
    \Bigg] \Op_L^{\rm lat}(1/a)
  \nonumber\\
  & & - \frac{\alpha_s}{4 \pi}\ \frac{1}{4}\ \Big[ d_2 - d^I \Big]
    \Op_P^{\rm lat}(1/a)
  \nonumber\\
  & & - \frac{\alpha_s}{4 \pi}\ \frac{1}{3}\ \Big[ w + w^I \Big]
    \Op_R^{\rm lat}(1/a)
  \nonumber\\
  & & + \frac{\alpha_s}{4 \pi}\ \frac{4}{3}\ 
    \Big[\ (1-c_{\rm sw})\ln\left(a^2\lambda^{2}\right) + V + V^I \Big]
    \Op_{SD}^{\rm lat}(1/a)
  \nonumber\\
  & & + \frac{\alpha_s}{4 \pi}\ \frac{-1}{2}\
    \Big[\ (1-c_{\rm sw})\ln\left(a^2\lambda^{2}\right) + V + V^I \Big]
    \Op_{LD}^{\rm lat}(1/a)
  \nonumber\\
  & & + \frac{\alpha_s}{4 \pi}\ \frac{1}{4}\ \Big[ U + U^I \Big]
    \Op_{PD}^{\rm lat}(1/a).
\end{eqnarray}
A result with $c_{sw}$=1 is used in Eq.~(\ref{eq:O_S_matching}).


\begin{table}
  \begin{center}
    \begin{tabular}{c|r}
      $c$         &    4.53 \\
      $d_1$       &    5.46 \\
      $d_2$       & $-$7.22 \\
      $d^I$       & $-$4.13 \\
      $e^{(R)}$   &    4.53 \\
      $f$         &   13.35 \\
      $f^I$       & $-$3.64 \\
      $v$         & $-$6.92 \\
      $v^I$       & $-$6.72 \\
      $w$         & $-$1.20 \\
      $w^I$       &    0.82 \\
      $J_1$       & $-$4.85 \\ 
      $U$         &    4.89 \\
      $U^I$       & $-$0.29 \\ 
      $V$         & $-$7.14 \\
      $V^I$       &    1.98 \\ 
      $u_0^{(2)}$(link) & $-\pi^2$ \\
      $u_0^{(2)}$($\kappa_c$)  & $-$8.00\\
    \end{tabular}
    \caption{Numerical values of parameters appearing in the
             one-loop lattice integrals.}
    \label{tab:PERT}
  \end{center}
\end{table}
\end{document}